# Ideal Four Wave Mixing Dynamics in a Nonlinear Schrödinger Equation Fibre System

**Anastasiia Sheveleva,[1] Ugo Andral,[1] Bertrand Kibler,[1] Pierre Colman,[1] John. M. Dudley,[2] Christophe Finot[1,*]**

[1] *Laboratoire Interdisciplinaire Carnot de Bourgogne, UMR 6303 CNRS-Université de Bourgogne-Franche-Comté, 9 avenue Alain Savary, BP 47870, 21078 Dijon Cedex, France*
[2] *Institut FEMTO-ST, Université Bourgogne Franche-Comté CNRS UMR 6174, Besançon, 25000, France.*
*\* Corresponding author: christophe.finot@u-bourgogne.fr*

**Near-ideal four wave mixing dynamics are observed in a nonlinear Schrödinger equation system using a new experimental technique associated with iterated sequential initial conditions in optical fiber. This novel approach mitigates against unwanted sideband generation and optical loss, extending the effective propagation distance by two orders of magnitude, allowing Kerr-driven coupling dynamics to be seen over 50 km of optical fiber using only one short fiber segment of 500 m. Our experiments reveal the full dynamical phase space topology in amplitude and phase, showing characteristic features of multiple Fermi-Pasta-Ulam recurrence cycles, stationary wave existence, and the system separatrix boundary. Experiments are shown to be in excellent quantitative agreement with numerical solutions of the canonical differential equation system describing the wave evolution.**

## 1. INTRODUCTION

The nonlinear Schrödinger equation (NLSE) is one of the seminal equations of science, describing wave evolution in a dispersive medium subject to an intensity-dependent nonlinear phase shift. It applies to many different domains including plasma physics, hydrodynamics, Bose-Einstein condensates, analog gravity, optical self-focussing and filamentation, and wave propagation in optical fiber [1].

The key physical process in the NLSE is nonlinear four wave mixing (FWM), which arises from the dispersion-mediated energy exchange between discrete evolving frequency components [2]. From a fundamental perspective, the essential features of FWM are most clearly seen in the degenerate case when a single frequency pump generates only two sidebands of upshifted and downshifted frequency. In this case the system is fully described by a reduced system of three coupled differential equations which fully captures the rich dynamical landscape [3]. This includes effects such as the initial phase of modulation instability, Fermi-Pasta-Ulam recurrence, stationary waves (fixed points), and a separatrix boundary between different dynamical regimes. However although this canonical FWM system has been the subject of a number of previous theoretical and numerical studies, it is notoriously difficult to implement in practice. Consequently the expected ideal FWM dynamics have only been partially or imperfectly seen in experiments to date.

Here, we address this shortcoming directly through a new experimental technique which has allowed us to characterize near-ideal FWM dynamics in a nonlinear Schrödinger equation system based on optical fibre propagation. In particular, we have developed a system where iterated initial conditions are sequentially injected into an optical fiber, extending the effective propagation distance by two orders of magnitude and mitigating against unwanted sideband generation and optical loss. As a result we are able to clearly follow the dynamical interactions between only four evolving frequency components over a distance of 50 km using only one fiber segment of only 500 m length. Our experiments reveal the full dynamical phase space topology, revealing characteristic features of multiple Fermi-Pasta-Ulam recurrence cycles, stationary wave existence, and the system separatrix boundary. We compare our experimental results with solutions of the canonical differential equation system describing the wave evolution, obtaining excellent agreement. This approach represents a significant improvement in both implementation and accuracy on previous approaches and moreover, it can be readily generalized to the study of any arbitrary number of interacting wave components. This represents a major advance in the development of experimental techniques in nonlinear fiber optics.

## 2. THEORETICAL BACKGROUND AND PRINCIPLE

We first review the theoretical description of ideal FWM dynamics in the NLSE, and use numerical simulations to illustrate the expected dynamical behavior. In an ideal single mode and loss-free fiber, the evolution of a slowly-varying electric field envelope $\psi(z,t)$ is governed by the nonlinear Schrödinger equation:

$$i \frac{\partial \psi}{\partial z} - \frac{\beta_2}{2} \frac{\partial^2 \psi}{\partial t^2} + \gamma |\psi|^2 \psi = 0, \qquad (1)$$

with $z$ being the propagation distance and $t$ the time in a reference frame traveling at the group velocity. The group-velocity dispersion is $\beta_2$ and the nonlinear Kerr coefficient is $\gamma$. We can write the NLSE in normalized form:

$$i\frac{\partial A}{\partial \xi} + \frac{1}{2}\frac{\partial^2 A}{\partial \tau^2} + |A|^2 A = 0, \quad (2)$$

Here, normalized propagation and co-moving time variables $\xi$ and $\tau$ are linked to the dimensional quantities in fibre optics by $\xi = z/L_{NL}$ and $\tau = t/\sqrt{|\beta_2|L_{NL}}$. The characteristic length scale is defined as: $L_{NL} = (\gamma P_0)^{-1}$ with $P_0$ a power variable which in our case corresponds to the average power of the injected signal. The normalized field is related to its dimensional equivalent $\psi(z,T)$ by $A(\xi,\tau) = \psi(z,T)/\sqrt{P_0}$. Note that this form of the NLSE that describes "focusing" dynamics is associated with a fibre dispersion parameter $\beta_2 < 0$.

We discuss the fundamental wave mixing processes in the NLSE by considering the injection of a modulated pump wave $A_0$ with two sidebands at frequencies $\pm\Omega$:

$$A(\xi,\tau) = A_0(\xi) + A_{-1}(\xi)\exp(i\Omega\tau) \\ + A_1(\xi)\exp(-i\Omega\tau), \quad (3)$$

Note that we omit the carrier frequency here, and the normalized frequency $\Omega$ is related to dimensional frequency $f_m$ in Hz by: $\Omega = 2\pi f_m \sqrt{|\beta_2|/\gamma P_0}$. In general, the injection of such a modulated signal in an optical fibre leads to the generation of multiple additional sidebands, but the ideal truncated FWM system which describes only pump and first sideband energy exchange with distance is described by only three coupled equations:

$$\begin{cases} -i\dfrac{dA_0}{d\xi} = \left(|A_0|^2 + 2|A_{-1}|^2 + 2|A_1|^2\right)A_0 \\ \qquad\qquad + 2\,A_{-1}A_1 A_0^* \\ -i\dfrac{dA_{-1}}{d\xi} + \dfrac{1}{2}\Omega^2 A_{-1} = \left(|A_{-1}|^2 + 2|A_0|^2 + 2|A_1|^2\right)A_{-1} \\ \qquad\qquad + A_1^* A_0^2 \\ -i\dfrac{dA_1}{d\xi} + \dfrac{1}{2}\Omega^2 A_1 = \left(|A_1|^2 + 2|A_0|^2 + 2|A_{-1}|^2\right)A_1 \\ \qquad\qquad + A_{-1}^* A_0^2 \end{cases}, \quad (4)$$

When $|A_{\pm 1}| \ll |A_0|$, amplification of the lateral sidebands can occur for $\Omega < 2$, with maximum gain at a frequency $\Omega = \Omega_0 = \sqrt{2}$. This is of course the same condition for maximum gain that is derived in the linear stability analysis of modulation instability [3]. Note also that even though this system describes degenerate FWM, the fact there are only three frequency components involved has led to it being described (somewhat confusingly) as a "three wave" system.

From a dynamical perspective, this system can be associated with the one-dimensional conservative Hamiltonian:

$$H = 2\eta(1-\eta)\cos\phi + (\Omega^2+1)\eta - \frac{3}{2}\eta^2. \quad (5)$$

with canonical conjugate variables $\eta = \eta(\xi)$ and $\phi = \phi(\xi)$ satisfying:

$$\frac{\partial \eta}{\partial \xi} = \frac{\partial H}{\partial \phi} \quad \text{and} \quad \frac{\partial \phi}{\partial \xi} = -\frac{\partial H}{\partial \eta}, \quad (6)$$

and where $\eta$ and $\phi$ are related to the amplitudes $A_k(\xi)$ and phases $\varphi_k(\xi)$ of the evolving sidebands ($k = 0, \pm 1$) by:

$$\begin{cases} \eta = \dfrac{|A_0|^2}{|A_0|^2 + |A_{-1}|^2 + |A_1|^2}, \\ \phi = \varphi_1 + \varphi_{-1} - 2\varphi_0 \end{cases} \quad (7)$$

Here, $\eta$ and $\phi$ have physical interpretations as the fraction of the total power in the central frequency component and the sideband-pump frequency component phase difference respectively. Tracing the dynamics in the ($\eta\cos\phi, \eta\sin\phi$) plane fully captures all the physics of this ideal system.

To illustrate the physics of this system [4], Fig. 1 shows modelling results for different initial conditions. The parameters chosen correspond to maximum gain with $\Omega = \Omega_0$, and we assume initially equal sideband intensities $A_1(0)=A_{-1}(0)$. In panels a and b we show results of numerical integration of the ideal FWM system in Eq. 4 (blue), results from simulation of the segmented approach (described below) which we use in our experiments (red), and results from numerical solution of the full multiwave interactions from the NLSE (yellow). In panels c we show the temporal evolution computed from Eq. 3 for the ideal FWM case.

We now discuss these results in detail. We first consider an initial value $\eta_0 = 0.9$ and in-phase initial components ($\phi_0 = 0$) shown in panels a1, b1, c1. The periodic evolution of $\eta$ in Panel 1a clearly shows reversible energy transfer from the central mode to the sidebands, associated with the expected Fermi-Pasta-Ulam recurrence [5, 6]. This recurrence is also seen in the corresponding temporal intensity profile (panel c1), and is reflected in the closed trajectories in the phase space portrait in panel b1. These orbits are localized on the right-hand side of the ideal FWM separatrix orbit (dashed black line, also computed from Eqs 4) which divides the phase space into two distinct regimes.

For $\eta_0 = 0.9$ but with $\phi_0 = \pi$, very different features are observed. This highlights the crucial role of the initial phase in this system. Although the power redistribution between modes remains periodic (panel a2), the amplitude variation is increased compared to the previous case. Moreover, the closed phase space trajectory (panel b2) is found on the opposite side of the separatrix. Physically, this is associated with modified temporal evolution as shown in panel c2, with a temporal phase-shift taking place in each recurrence cycle associated with longitudinal period doubling [6, 7].

.

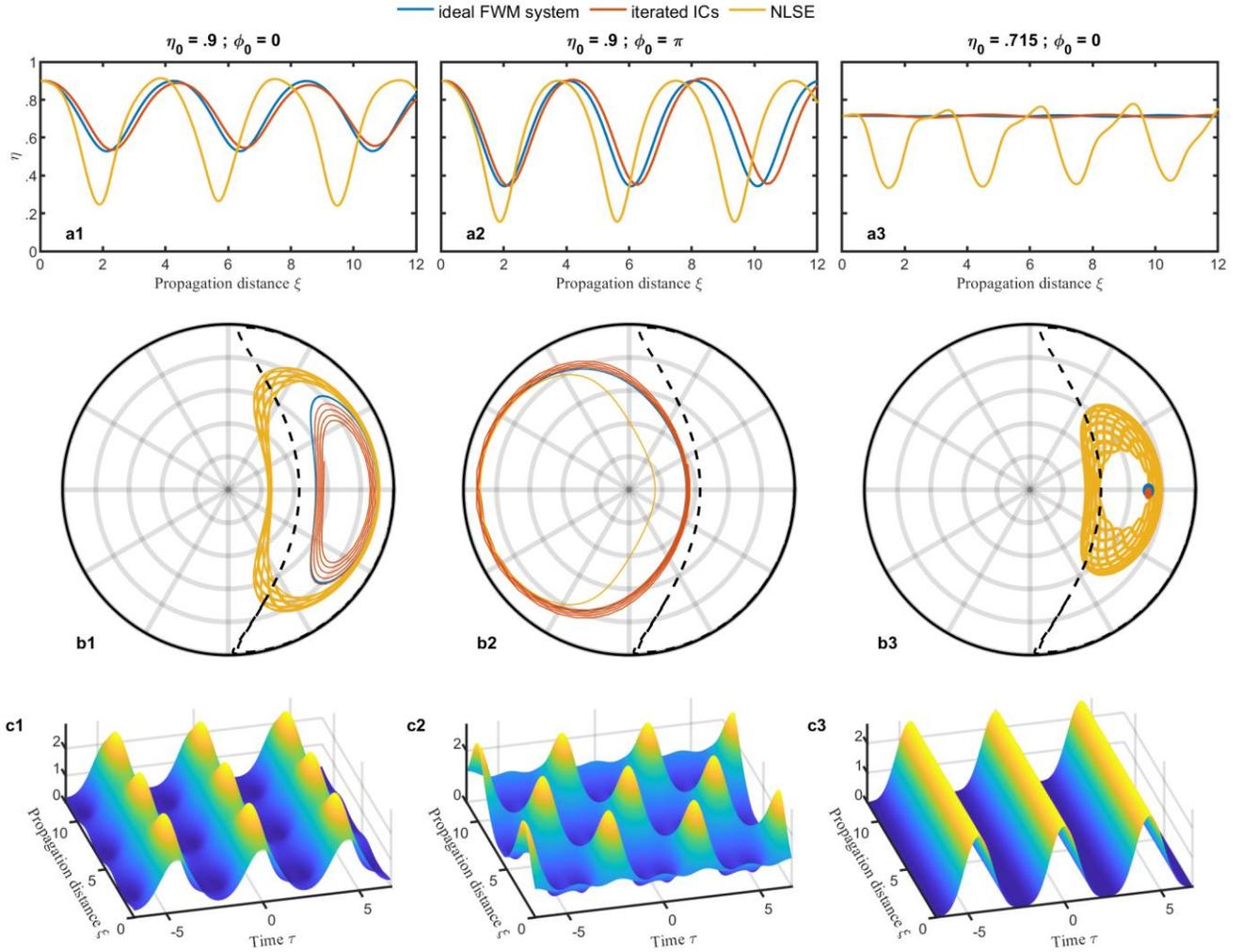

**Fig. 1.** Evolution properties of an strong pump and two lateral sidebands spaced by $\Omega = \Omega_0 = \sqrt{2}$. Panels 1, 2 and 3 show respectively results for different initial conditions: $\eta_0 = 0.9$ and $\phi_0 = 0$, $\eta_0 = 0.9$ and $\phi_0 = \pi$, $\eta_0 = 0.715$ and $\phi_0 = 0$. Subpanels (a) show evolution of $\eta$ for the ideal FWM system (blue), a segmented approach (red) and the full NLSE (yellow). (b). Corresponding hase-space portraits. (c) Corresponding evolution of the temporal intensity profile for the ideal FWM system.

As a final example, panels a3-c3 show the dynamics observed for $\eta_0 = 0.715$ and with $\phi_0 = \pi$. This leads to a near-stationary solution with very low amplitude variation in the side band ratio $\eta$ (panel a3) and in the temporal intensity profile (panel c3). The phase space trajectory in this case is a closed orbit of very small effective radius, close to a fixed point of the system (panel b3).

Panels a1-a3 in Figure 1 also show the evolution of $\eta$ computed from the numerical simulation of the full NLSE including multiple sideband generation (yellow). These results are clearly very different from the truncated ideal FWM system, with significantly more depletion of the central frequency component as additional sidebands are generated [8]. These differences are also very apparent when comparing the ideal FWM (blue) and NLSE (yellow) orbits in the phase space portraits in panels b1-b3. In fact, we see that the NLSE orbits actually cross the separatrix associated with the ideal FWM system (this is in fact expected here given that the NLSE separatrix for this case is associated with the Akhmediev breather which possesses an infinite number of sidebands.) Moreover, for $\eta_0 = 0.715$, the stationary solution is clearly not recovered and the orbit is much more complex than a fixed point.

These examples clearly show the difficulties in observing the canonical dynamics of ideal FWM in a NLSE system – the generation of additional sidebands in the full NLSE leads to major quantitative disagreement. However, a segmented approach to propagation with reinjection of power-adjusted initial conditions allows us to overcome this limitation, and to develop a practical system that yields close to ideal FWM dynamics. The principle here is to replace a single long segment of fiber by a concatenation of segments that are sufficiently short such that additional sidebands cannot reach a significant level. Moreover, between sequential segments we cancel spectral components outside the four principle modes, and we use amplification to restore the same average power. Results illustrating this segmented approach are shown as the red lines in Fig. 1, and clearly show how this approach yields excellent agreement with the ideal FWM model: all the main features previously discussed are now quantitatively reproduced. Note that for these results, we use fiber segment lengths of $\xi_L = 0.12$, a choice which is motivated by our experiments described in Section 4.

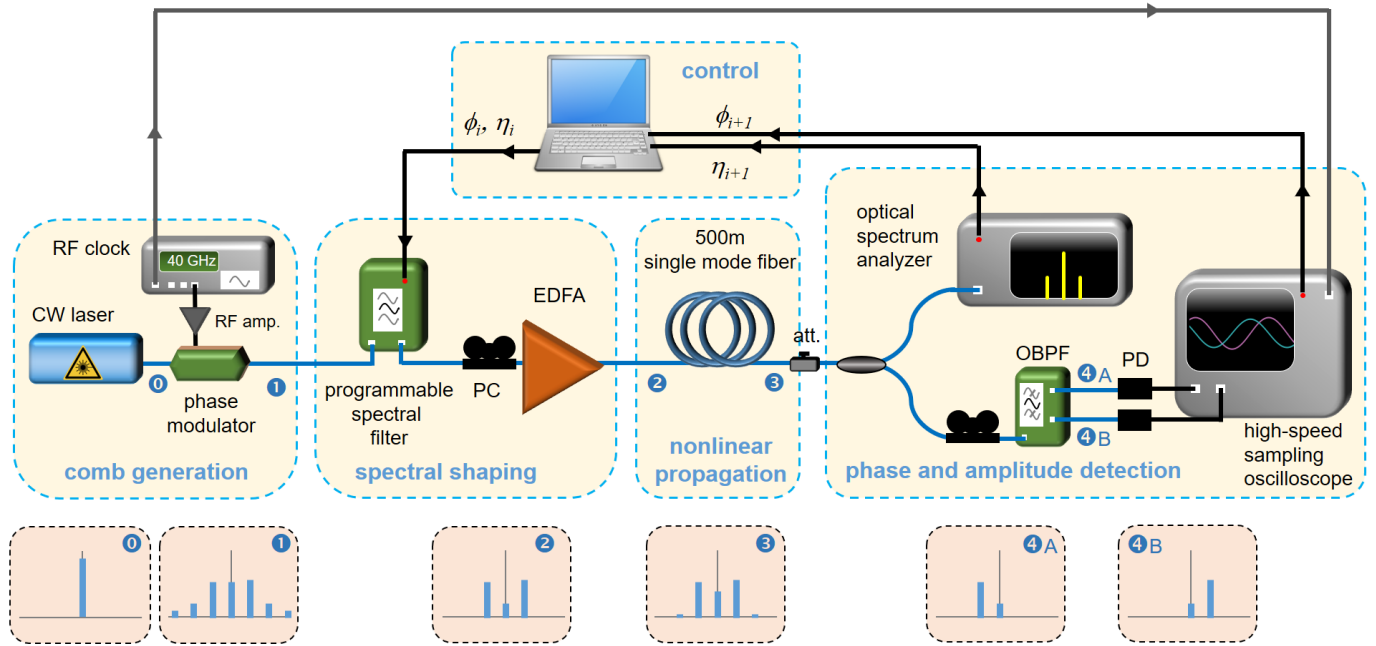

**Fig. 2.** Experimental setup. The insets 0-4 in the bottom panels represent the optical spectrum generated at different stages of the experiment.

## 3. EXPERIMENTAL SETUP

Attempts to measure the complex longitudinal NLSE wave mixing dynamics in optical fibre have been previously reported using various methods such as destructive cut-back measurements [8], distributed optical time domain reflectometry [9, 10] or evolution in a recirculating loop [11, 12]. However, deviation between experiments and ideal FWM dynamics becomes significant very quickly in these cases, and is dramatically impacted by even small amounts of distributed loss or gain [5, 9, 13].

Our experimental setup is shown in Fig. 2 and is made of commercially-available telecommunications components. First, a laser operating at 1550 nm emits a continuous wave (CW). A phase modulator driven by a 40 GHz RF sinusoidal modulation converts the monochromatic laser spectrum into a set of equally spaced spectral lines [14]. The resulting symmetrical comb is then processed using a programmable filter (waveshaper device based on liquid crystal on silicon [15]) that simultaneously implements several operations: elimination of unwanted spectral components, and the precise adjustment of the ratio $\eta$ between the central and lateral components as well as their relative phase $\phi$. Special care has been devoted to ensure that no unwanted phase/intensity coupling occurs during this shaping. The tailored three-component signal with the target $\eta_i$ and $\phi_i$ is then amplified by an erbium-doped fiber amplifier that can deliver a tunable power. The amplifier runs in a power controlled mode so that the average power does not depend on the input spectral properties such that the system can be considered as quasi-conservative.

Nonlinear propagation takes place in single mode optical fiber with dispersion and nonlinear parameters being respectively -7.6 $ps^2.m^{-1}$ and 1.7 $W^{-1}.m^{-1}$. The fiber length is 500 m, with this length selected as a tradeoff between the sensitivity of the detection stage of our setup and the appearance of Brillouin scattering: with 500-m of this fiber, the changes experienced by the optical field are significant enough to be conveniently detected and we have checked the absence of Brillouin backscattering for the range of powers under investigation. We can therefore work with CW signals without having to involve additional strategies of temporal pulse carving and associated synchronization. In order to limit polarization mode dispersion, the input state of polarization is optimized using polarization controllers.

The output signal is then attenuated and split into two in order to record both its spectral phase and intensity. An optical spectrum analyzer (OSA, resolution 0.1 nm) provides directly the ratio $\eta_{i+1}$. The spectral phase offset $\phi_{i+1}$ is retrieved from the temporal delay between the central and lateral sidebands as measured with a high-speed sampling oscilloscope. The experimentally measured values can then be imprinted as new input values and the process can be iterated at will without any accumulation of deleterious amplified spontaneous emission and without any growth of unwanted spectral sidebands or noise. Potentially unlimited propagation can therefore be emulated, similarly to methods that have for example been implemented in the field of hydrodynamics [16].

## 4. EXPERIMENTAL RESULTS

### A. Phase space and longitudinal reconstruction of the dynamics

We first study the dynamics of the system at maximum gain with $\Omega = \Omega_0 = \sqrt{2}$, i.e. for $P_{in}$ = 21.5 dBm. In terms of normalized units, $L_{NL}$ = 4.1 km and the 500m length of our fiber segment corresponds to a normalized length $\xi_L$ = 0.12, similar to the one used in the discussion of section 2C. The experimental phase space portraits obtained for different initial values $\eta_0$ and $\phi_0$ are shown in Fig. 3(a) with the orbits shown as circles connected by dotted lines. Note that in these experiments, we checked that the energy contained in the unwanted sidebands located at $\pm 2 f_m$ always remained well below 3% of the total energy of the signal.

For each value of $\eta_0$, we examined the dynamics at two values of initial phase: $\phi_0$ = 0 and $\phi_0$ = $\pi$ which yielded trajectories on the right and the left of the separatrix as expected from the results in Section 2. The dynamics were measured over 25 km (i.e. 50 iterations) and the results yield immediate insight into the phase space topology. The experimental orbits are seen to be in very good agreement with the predictions from the ideal system described in section 2 which are shown as thick solid lines. Indeed, many fundamental features of the ideal FWM dynamics can be seen from these results. Specifically, we clearly confirm the importance of the separatrix dividing the phase plane into two well-defined regions, with the measurements for $\eta_0$ = 0.95 in particular providing a very clear indication of its location. We also see that the different experimental trajectories are nearly closed orbits and do not intersect. The slight discrepancies between experiment and prediction here are attributed to the accumulation of small errors in the phase/intensity measurements and residual depolarization effects not included in our scalar model.

Significantly, with complete experimental knowledge of the spectral phase and intensity of the three interacting frequency components of the evolving field, it is straightforward to fully reconstruct the evolving intensity profiles in the temporal domain. Over a propagation distance of 50 km (a normalized distance $\xi$ = 12, 100 iterations), Fig. 3(b) shows these results for an initial value of $\eta_0$ = 0.9 and both initial values of phase $\phi_0$ = 0 and $\phi_0$ = $\pi$, plotted beneath the corresponding orbits on the right and left side of the phase space plot. These results show the expected recurrence dynamics as seen in Section 2, and for the case of $\phi_0 = \pi$, also highlight the evolution phase shift of half temporal period, leading, as expected by theory to a period doubling [7]. Note that these double-periodic solutions have been the subject of particular recent interest [9, 17].

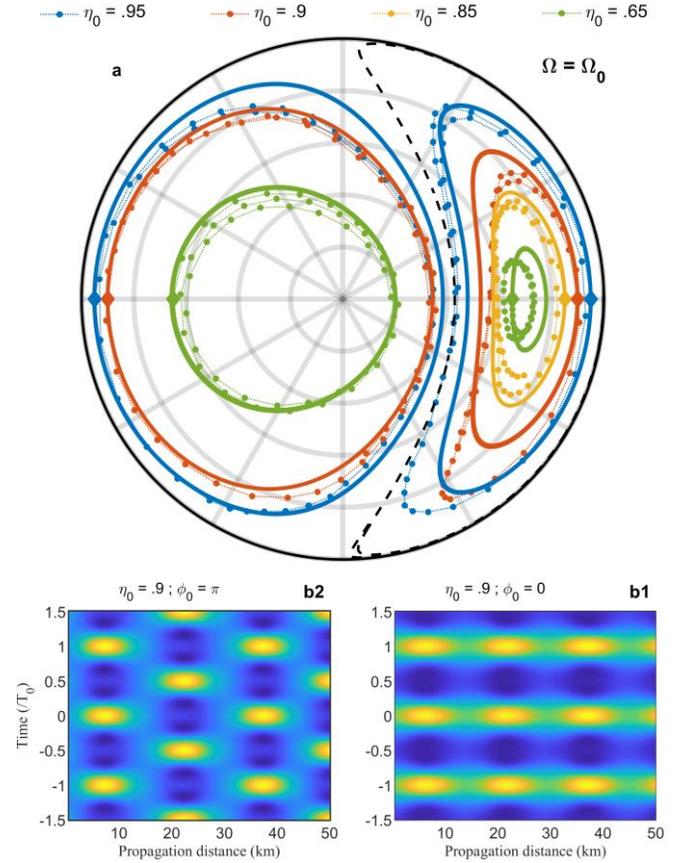

**Fig. 3.** Experimental results for $\Omega = \Omega_0 = \sqrt{2}$. (a) Phase space portraits for initial values of $\eta_0$ of 0.65, 0.85, 0.90 and 0.95 (green, yellow, red and blue lines respectively). Results are plotted for an initial phase offset $\phi_0$ of 0 or $\pi$. which appear respectively on the right and left side of the separatrix (dashed black line). The experimental results over 50 iterations (circles joined with dotted lines) are compared with the theoretical solution of system (4) (solid thick line). (b) Longitudinal evolution of the temporal intensity profiles reconstructed from the experimental spectral measurements for $\eta_0$ of 0.9 and phases of 0 and $\pi$ as indicated.

### B. Influence of the gain

Tuning the input power, we can also explore the modulationally unstable dynamics for higher values of gain. Phase space portraits obtained for an average value of 23.7 dBm leading to $\Omega$ = 1.1 are plotted in Fig. 4. Once again, the experimental results are in good agreement with the theoretical predictions. And when comparing with Fig. 3(a), we note how the dynamics at higher gain are associated with the change of the shape of the trajectories and the displacement of the separatrix.

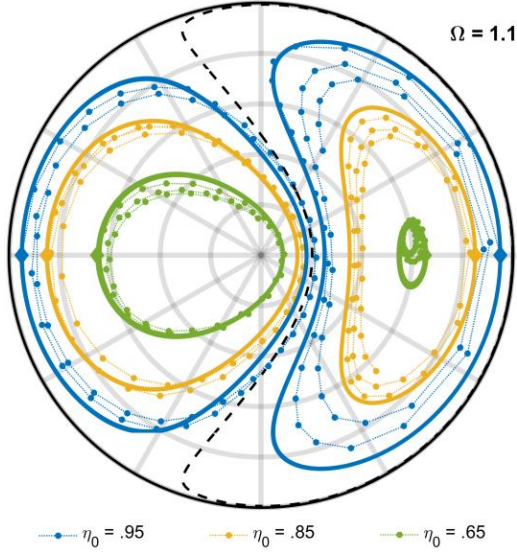

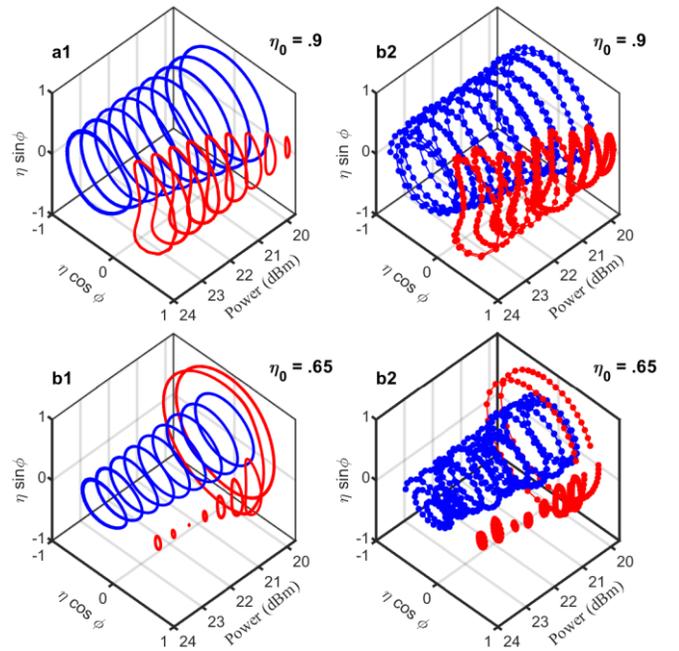

**Fig. 4.** Experimental phase portraits obtained at an average power $P_{in}$ = 23.7 dBm and for initial values of $\eta_0$ of 0.65, 0.85 and 0.95 (green, yellow and blue lines respectively). Results are plotted for an initial phase offset $\phi_0$ of 0 or $\pi$, which appear respectively on the right and left side of the separatrix (dashed black line). The experimental results accumulated over 50 iterations (circles joined with dotted lines) are compared with the theoretical solution of (4) (solid thick line).

**Fig. 5.** Influence of the input average power on the phase space portraits obtained for initial values of $\eta_0$ of 0.9, and 0.65 (panels a and b respectively) and initial phase offset of 0 or $\pi$. The theoretical predictions (panels 1) are compared with the experimental results (panels 2).

A more exhaustive study of the influence of the average power for a fixed value of $\eta_0$ = 0.65 and $\eta_0$ = 0.9 and phase offset $\phi_0$ of 0 and $\pi$ is shown in Fig. 5. Average powers between 19.7 and 23.7 dBm were tested, leading in terms of normalized frequency $\Omega$ of a range between 1.74 and 1.1.

The measurements of the instability process achieved at $\eta_0$ =0.9 (panels a) confirm that with increasing powers the separatrix progressively shifts: the intersection point between the separatrix and the horizontal axis $\phi$ = 0 continuously decreases. Consequently, the phase space available for the evolution of initial conditions $\phi_0$= 0 gets larger and larger whereas initial conditions $\phi_0 = \pi$ evolve in more and more restricted areas.

Further measurements at $\eta_0$ =0.65 (panels b) are also of interest, especially for $\phi_0$ = 0. Indeed, we note that for the lowest powers (19.7 and 20.2 dBm), the initial condition $\eta_0$ =.65 and $\phi_0$ = 0 leads to orbits that are on the left side of the separatrix. For these powers, the trajectory obtained for $\phi_0 = \pi$ is therefore surrounded within the trajectory for $\phi_0$ = 0. When increasing the power, the separatrix is crossed and each initial condition evolves on a different side of the phase plane. For powers between 20.7 and 22.7 dBm, the orbits get smaller and smaller up to the stage where they reach a fixed point for 22.7 dBm. When further increasing the average power, the orbit becomes increasingly open.

### C. Observation of a fixed point

Finally, we investigate in more detail the properties observed at one of the two fixed points of the phase plane. For $\eta_0$ = 0.65 and $\phi_0$ = 0 and $P_{in}$ = 22.7 dBm ($\Omega$ = 1.23), the longitudinal evolution of the temporal intensity reconstructed from the spectral measurements is plotted in Fig. 6(a) over 50 km. We clearly see in this case that the temporal profile is invariant with propagation. Panel (b) explicitly compares these results with temporal measurements made with a picosecond-resolution optical sampling oscilloscope and the agreement is such that they cannot be visually distinguished. Moreover, both experimental profiles agree with the expected temporal profile computed from Eq. 3 which in the case of a fixed point simply consists of a stationary temporal profile formed from three wave interference.

More generally, a fixed point of the ideal FWM system is predicted to exist at any value of gain for a particular choice of sideband ratio. Expressed in terms of normalized frequency $\Omega$, the dependence is given by [4]:

$$\eta_e = \frac{3+\Omega^2}{7}, \qquad (8)$$

and this can be readily tested experimentally. Specifically, for different input powers, the waveshaper is used to experimentally determine the value of $\eta_e$ associated with the fixed point, and the results are shown in Fig. 3(c). The agreement between experimental results (blue) and the prediction of Eq. 8 is excellent.

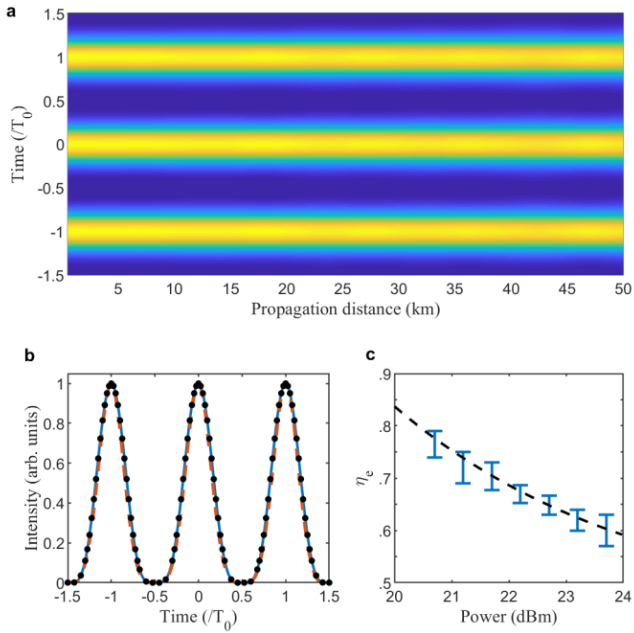

**Fig. 6.** (a) Longitudinal evolution of the temporal intensity profile reconstructed from the spectral properties of the signal obtained for an average power of 22.7 dBm and initial conditions $\eta_0 = 0.65$ and $\phi_0 = 0$. (b) Comparison of the temporal intensity profile reconstructed from spectral measurements (blue line) and directly measured with an optical sampling oscilloscope (red dashed line). Results from theoretical predictions are also displayed with black circles. (c) Evolution of the value of $\eta_0$ leading to a fixed point. The experimental data (blue) are compared with the analytical prediction (dashed black line, Eq. (8)).

## 5. CONCLUSION AND OUTLOOK

Optical systems are well known to provide flexible testbeds with which to study the physics of diverse nonlinear systems, and the results here show the success of a new experimental approach allowing the dynamics of ideal four wave mixing to be fully explored. The use of iterated initial conditions in a short fiber segment mitigates against effects of loss, inelastic scattering and high-order sideband generation, allowing clear observation of the predicted evolution dynamics of the FWM system. Amongst the dynamics that are seen experimentally are Fermi-Pasta-Ulam recurrence, qualitatively different evolution on either side of a separatrix, and the existence of system fixed points.

It is important to stress the adaptability of this experimental technique. Although we have focused on spectrally-symmetric initial conditions, experiments can be readily adapted to asymmetric initial sidebands [4] and to evolution in fibers with normal dispersion. This opens up the possibility to study even more complex nonlinear wave mixing [17, 18]. Vector instability processes in birefringent fibers offer additional perspectives [19], and the principle of the technique can be extended into the spatial domain using spatial light modulators and phase-intensity characterization [20]. We also anticipate extension to other branches of nonlinear physics where discrete wave mixing plays a central role in the system evolution [21].

**Funding.** OPTIMAL (ANR-20-CE30-0004); EIPHI-BFC (ANR-17-EURE-0002); I-SITE BFC (ANR-15-IDEX-0003); Région Bourgogne-Franche-Comté; Institut Universitaire de France.

**Acknowledgments.** The numerical simulations used the HPC resources of DNUM CCUB (Centre de Calcul de l'Université de Bourgogne). The authors also thank GDR Elios (GDR 2080).

**Disclosures.** The authors declare no conflicts of interest.

**Data availability.** The data that support the findings of this study are available from the corresponding author, CF, upon reasonable request.

## References

1. C. Sulem and P.-L. Sulem, *The nonlinear Schrödinger equation: self-focusing and wave collapse* (Springer Science & Business Media, 2007), Vol. 139.
2. R. W. Boyd, *Nonlinear optics* (Academic press, 2003).
3. G. P. Agrawal, *Nonlinear Fiber Optics, Fourth Edition* (Academic Press, San Francisco, CA, 2006).
4. G. Cappellini and S. Trillo, "Third-order three-wave mixing in single-mode fibers: exact solutions and spatial instability effects," J. Opt. Soc. Am. B **8**, 824-838 (1991).
5. G. Van Simaeys, P. Emplit, and M. Haelterman, "Experimental study of the reversible behavior of modulational instability in optical fibers," J. Opt. Soc. Am. B **19**, 477-486 (2002).
6. A. Mussot, C. Naveau, M. Conforti, A. Kudlinski, F. Copie, P. Szriftgiser, and S. Trillo, "Fibre multi-wave mixing combs reveal the broken symmetry of Fermi–Pasta–Ulam recurrence," Nat. Photon. **12**, 303-308 (2018).
7. S. Trillo and S. Wabnitz, "Dynamics of the nonlinear modulational instability in optical fibers," Opt. Lett. **16**, 986-988 (1991).
8. K. Hammani, B. Wetzel, B. Kibler, J. Fatome, C. Finot, G. Millot, N. Akhmediev, and J. M. Dudley, "Spectral dynamics of modulation instability described using Akhmediev breather theory," Opt. Lett. **36**, 2140-2142 (2011).
9. C. Naveau, G. Vanderhaegen, P. Szriftgiser, G. Martinelli, M. Droques, A. Kudlinski, M. Conforti, S. Trillo, N. Akhmediev, and A. Mussot, "Heterodyne Optical Time Domain Reflectometer Combined With Active Loss Compensation: A Practical Tool for Investigating Fermi Pasta Ulam Recurrence Process and Breathers Dynamics in Optical Fibers," Front. Phys. **9** 637812 (2021).
10. X. Hu, W. Chen, Y. Lu, Z. Yu, M. Chen, and Z. Meng, "Distributed Measurement of Fermi–Pasta–Ulam Recurrence in Optical Fibers," IEEE Photon. Technol. Lett. **30**, 47-50 (2018).
11. J.-W. Goossens, H. Hafermann, and Y. Jaouën, "Experimental realization of Fermi-Pasta-Ulam-Tsingou recurrence in a long-haul optical fiber transmission system," Sci. Rep. **9**, 18467 (2019).
12. A. E. Kraych, P. Suret, G. A. El, and S. Randoux, "Nonlinear Evolution of the Locally Induced Modulational Instability in Fiber Optics," Phys. Rev. Lett. **122**, 054101 (2019).
13. O. Kimmoun, H. C. Hsu, H. Branger, M. S. Li, Y. Y. Chen, C. Kharif, M. Onorato, E. J. R. Kelleher, B. Kibler, N. Akhmediev, and A. Chabchoub, "Modulation Instability and Phase-Shifted Fermi-Pasta-Ulam Recurrence," Sci. Rep. **6**, 28516 (2016).
14. K. Hammani, J. Fatome, and C. Finot, "Applications of sinusoidal phase modulation in temporal optics to highlight some properties of the Fourier transform," Eur. J. Phys. **40**, 055301 (2019).
15. A. M. Clarke, D. G. Williams, M. A. F. Roelens, and B. J. Eggleton, "Reconfigurable Optical Pulse Generator Employing a Fourier-Domain Programmable Optical Processor," J. Light. Technol. **28**, 97-103 (2010).

16. A. Chabchoub, N. Hoffmann, M. Onorato, and N. Akhmediev, "Super Rogue Waves: Observation of a Higher-Order Breather in Water Waves," Phys. Rev. X **2**, 011015 (2012).
17. A. Barthelemy and R. De La Fuente, "Unusual modulation instability in fibers with normal and anomalous dispersions," Opt. Commun. **73**, 409-412 (1989).
18. S. Trillo, S. Wabnitz, and T. Kennedy, "Nonlinear dynamics of dual-frequency-pumped multiwave mixing in optical fibers," Phys. Rev. A **50**, 1732-1747 (1994).
19. S. G. Murdoch, R. Leonhardt, and J. D. Harvey, "Nonlinear dynamics of polarization modulation instability in optical fiber," J. Opt. Soc. Am. B **14**, 3403-3411 (1997).
20. D. Pierangeli, M. Flammini, L. Zhang, G. Marcucci, A. J. Agranat, P. G. Grinevich, P. M. Santini, C. Conti, and E. DelRe, "Observation of Fermi-Pasta-Ulam-Tsingou Recurrence and Its Exact Dynamics," Phys. Rev. X **8**, 041017 (2018).
21. J. R. Thompson and R. Roy, "Statistical fluctuations in multiple four-wave mixing in a single-mode optical fiber," Phys. Rev. A **44**, 7605-7614 (1991).